\documentclass{article}

     \usepackage[final]{neurips_2019}

\usepackage[utf8]{inputenc} 
\usepackage[T1]{fontenc}    
\usepackage{hyperref}       
\usepackage{url}            
\usepackage{booktabs}       
\usepackage{amsfonts}       
\usepackage{nicefrac}       
\usepackage{microtype}      
\usepackage{graphicx}
\usepackage{enumitem}
\usepackage{array,multirow}
\usepackage{media9}
\usepackage{hyperref}
\usepackage{amsmath}
\usepackage{appendix}

\hypersetup{
    colorlinks=true,
    linkcolor=black,
    filecolor=black,      
    citecolor=black,
    urlcolor=cyan
}

\title{GAN-enhanced Conditional Echocardiogram Generation}

\author{%
Amir H. Abdi \\
\texttt{amirabdi@ece.ubc.ca} \And
Teresa Tsang \\
\texttt{t.tsang@ubc.ca} \And
Purang Abolmaesumi  \\
\texttt{purang@ece.ubc.ca} \AND \\
 University of British Columbia, Vancouver, Canada
}

\begin{document}

\maketitle

\begin{abstract}
  Echocardiography (echo) is a common means of evaluating cardiac conditions.
Due to the label scarcity,
semi-supervised paradigms in automated echo analysis are getting traction.
One of the most sought-after problems in echo is the segmentation of cardiac structures (\emph{e.g.} chambers).
Accordingly, we propose an echocardiogram generation approach using  generative adversarial networks with a conditional patch-based discriminator.
In this work, we validate the feasibility of GAN-enhanced echo generation with different conditions (segmentation masks), namely, the left ventricle, ventricular myocardium, and atrium.
Results show that the proposed adversarial algorithm can generate high-quality echo frames whose cardiac structures match  the given segmentation masks.
This method is expected to facilitate the training of other machine learning models in a semi-supervised fashion as suggested in similar researches.
\end{abstract}

\section{Introduction}

Echocardiography (echo) is the main means of cardiac evaluation and diagnosis of heart conditions. 
Thanks to the large echo datasets,
deep learning solutions are vastly proposed to automate cardiac function analysis.
Automatic segmentation of echo structures is the basis for many cardiac indices, such as the ejection fraction, and plays a vital role in evaluation and diagnosis of cardiac conditions.


Due to the label scarcity and the unavailability of annotations for non-key echo frames, and in the wake of semi-supervised learning, researchers have proposed  hybrid solutions for the segmentation problem  to leverage the large quantity of unlabeled echo frames.
In a recent work, the T-L network architecture~\cite{girdhar2016learning} was used to generate a blurry estimation of the echo image based on a given segmentation mask. This generative model was then used as a regularization term for the objective function of segmentation and was shown to improve the overall accuracy of the solution~\cite{isbi2019}.

In this work, we are proposing a generative solution to  synthesize echo frames using the patch-based conditional generative adversarial network (Patch-cGAN).
In this approach, the segmentation masks are used as conditions for the image synthesis.
The public dataset of CAMUS is used for training and testing~\cite{CAMUS}, and the feasibility of the proposed method is subjectively evaluated.

\section{Method}

In this work, a method is proposed to generate echocardiogram frames based on the segmentation masks of one or multiple cardiac chambers  and the left ventricular myocardium. 
As per the common architecture of GANs, the proposed learning algorithm simultaneously optimizes two models, namely, the generator ($G$) and the adversarial discriminator ($D$), both of which are convolutional networks.



While GANs commonly follow a stochastic paradigm  for data generation, here, we take a deterministic path with no randomness neither during  training nor inference. 
This training is carried out in a paired fashion and
there exists only a single target echo frame for every segmentation mask in the CAMUS dataset. Therefore, and as suggested by Isola \emph{et al.}~\cite{pix2pix}, any imposed randomness is expected to be ignored by the model.
Consequently, the generator is trained to learn a deterministic mapping between the two distributions.

Here, a patch-based adversarial discriminator independently validates the realness of  each image patch.
This approach naively assumes independence between pixels of distant patches by modeling the image as a Markov random field~\cite{pix2pix}. 

As opposed to the binary cross entropy (BCE) criterion in the conventional cGAN,
here, the least squared error (LSE)  was used as the adversarial term. 
This choice forces the discriminator to yield designated values  for the fake and the real  samples. 
The LSE pushes the generated samples closer to the real data distribution and is shown to outperform the conventional softmax BCE~\cite{LSGAN2,GANSuggestions}:
\begin{equation}
    \mathcal{L}_{cGAN}(G,D) = \mathbb{E}_{x, y} \big[~ (1 - D(y, x))^2 ~\big] +
    \mathbb{E}_{x} \big[~ D(y, G(y)) ^2 ~\big] 
    ~,
\end{equation}
\begin{equation}
    G^* =  \texttt{arg}~ \underset{G}{\mathrm{min}}~ \underset{D}{\mathrm{max}} 
    ~ \lambda \mathcal{L}_{cGAN}(G, D) + \mathcal{L}_{Recon}(x, G(y)) ~.
\label{eq:main}
\end{equation}
Here, $y$ and $x$ denote the condition and the target echo image, respectively.
In Eq.~\ref{eq:main},  $\lambda$ is a constant that weighs the conditional adversarial term against the reconstruction term.
Pixel-wise mean average error is used as the reconstruction criterion in $\mathcal{L}_{Recon}(x, G(y))$.

\subsection{Neural Architectures}

The design of neural architectures for the generator and the discriminator are inspired by the DCGAN~\cite{DCGAN}.
In summary, the generator is a  UNet-like network,  composed of 7 convolutional and 7 deconvolutional layers without skip connections.
The discriminator has 5 convolutional layers and operates on the concatenation of the image and its condition.
The deconvolutions were implemented as transposed convolutions.
Outputs of all convolutional and deconvolutional layers  were batch-normalized, except for the last layers of $D$ and $G$, and followed by the Leaky ReLU activations. 
Both architectures end with a convolutional layer of stride 1 to avoid  pixelated images.
Kernels of size  $4 \times 4$ were  used for the most part.
More details of the architectures are available in the  released code.

\section{Experiments}

For the study to be reproducible, all experiments were carried out on  the public dataset of CAMUS~\cite{CAMUS}.
This dataset contains  450 patient studies. Each  study  includes an echo cine of the apical four-chamber view  (4CH) 
with segmentation masks of the end-systolic (ES) and end-diastolic (ED) frames.
The masks include segmentations of the left ventricle,  left ventricular myocardium, and atrium.

\begin{figure}
\centering
\includegraphics[trim={0 0 0 0},clip,width=0.9\textwidth]{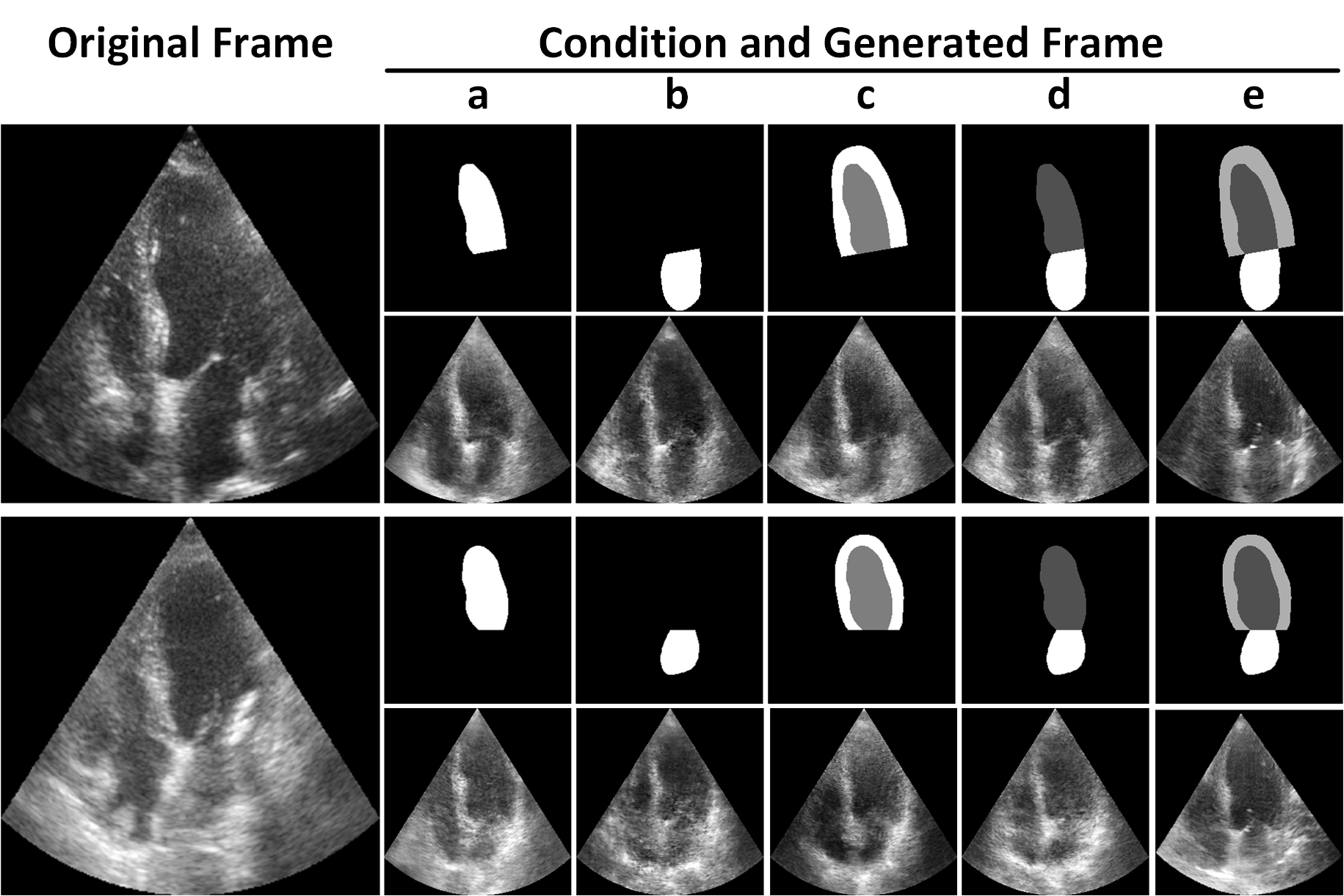}
\caption{Generated end-diastolic frames conditioned on  the segmentation masks and  their corresponding original (real) frames; a) Left ventricle b) Left atrium, c) Left ventricle and ventricular myocardium, d) Left ventricle and atrium, e) Left ventricle, ventricular myocardium, and atrium.} 
\label{fig:results_main}
\end{figure}

As shown in Figure~\ref{fig:results_main}, five experiments were conducted where the following segmentation masks were used as the condition:
1) Left ventricle,
2) Left atrium,
3) Left ventricle and ventricular myocardium,
4) Left ventricle and atrium,
5) Left ventricle, ventricular myocardium, and  atrium.
Among the 450 samples, 22 cases were randomly, but consistently, set aside for testing in all experiments.



\subsection{Training and Hyper-parameters}

The generator and the discriminator
were both trained with the Adam optimizer with the learning rates of 0.00013 and 0.00015, respectively.
Batch size was set to 8.
Size of patches for the patch-based discriminator were set to $16 \times 16$ pixels.
The weight of the adversarial term was set to $\lambda=0.01$.
Models were trained for 100k iterations.

No pre-processing was applied on the samples of the dataset except for the resizing of samples to $256 \times 256$ and normalizing the intensity of the gray-scale images to the range of $[0-1]$.
Similarly, no data augmentation were used during training.
The original values of the segmentation masks provided by the CAMUS dataset   were utilized as the conditions for the generator, where 0 is the background, 1 denotes the ventricle, 2 denotes the myocard, and 3 is the atrium.

The models were implemented with the TensorFlow deep learning library using the Keras API. Our code, scripts and configurations to reproduce the five experiments are publicly shared here: \url{https://github.com/amir-abdi/echo-generation}.

\subsection{Results and Discussion}

Results of evaluating the five experiments on the test set are presented in Figure~\ref{fig:results_main}.
Note that, during training,  the models have not seen the demonstrated target images of the test set nor their conditions.

As demonstrated, 
the generative models have learned a mapping from the segmentation masks to their corresponding cardiac structures, \emph{i.e.}, chambers and myocardium.
As a result, the trained models can generate apical four-chamber views of the heart, irrespective of the size of the segmentation mask and the quality of the original echo frame  (see Appendix~\ref{appendix:samples}).
A video demo of the echo generation from manually drawn left ventricle segmentation is available here: \url{https://youtu.be/9rfaL2uxkyc}.

We find it necessary to highlight that the results in columns a, b, c, d, and e of Figure~\ref{fig:results_main}  are generated by different models trained on different conditions; 
consequently, some levels of disagreement between their generations are expected.
At the same time, 
all models are trained on the same echo frames; thus,
generated samples are to some extent similar, specially in the affinity of  the given segmentation masks.
For example, 
 the models trained on the conditions containing the left ventricular segmentation masks (experiments a, c, d, and e) result in  relatively similar left ventricular regions.


As for the limitations of this work, we  only considered the end-diastolic (ED) frames to train the models; therefore, it is expected of the generator to only have learned this particular distribution.
Moreover, our domain of work was limited to the 4CH echo view.
Lastly, as for this preliminary feasibility study, 
no quantitative studies were conducted.
As a reasonable next step, we are aiming to generate the entire echo cine conditioned on the ED and ES frames. 


While we wished to design a probabilistic generative approach to address the one-to-many reality of echo generation conditioned on the segmentation mask, the unavailability of data held us back from this goal.
Consequently, 
we would like to remind our readers that no generative model can learn beyond its training data.
Therefore, any similarities between the generated samples and their corresponding original frames  should be associated with the limited variations in the ED echo frames of the limited samples of the CAMUS dataset  (see second row of Figure~\ref{fig:results1} in  Appendix A). 

\section*{Acknowledgement}
The authors would like to thank Mohammad H. Jafari (mohammadj@ece.ubc.ca) for inspiring us with his semi-supervised approach to left ventricle segmentation.
This work was conducted thanks to the funding from the Canadian Institutes of Health Research (CIHR). 

\bibliography{mybib}
\bibliographystyle{plain}

\newpage

\appendix
\appendixpage

\section{More Results}
\label{appendix:samples}

\begin{figure}[!h]
\includegraphics[trim={0 0 0 0},clip ,width=1.0\textwidth]{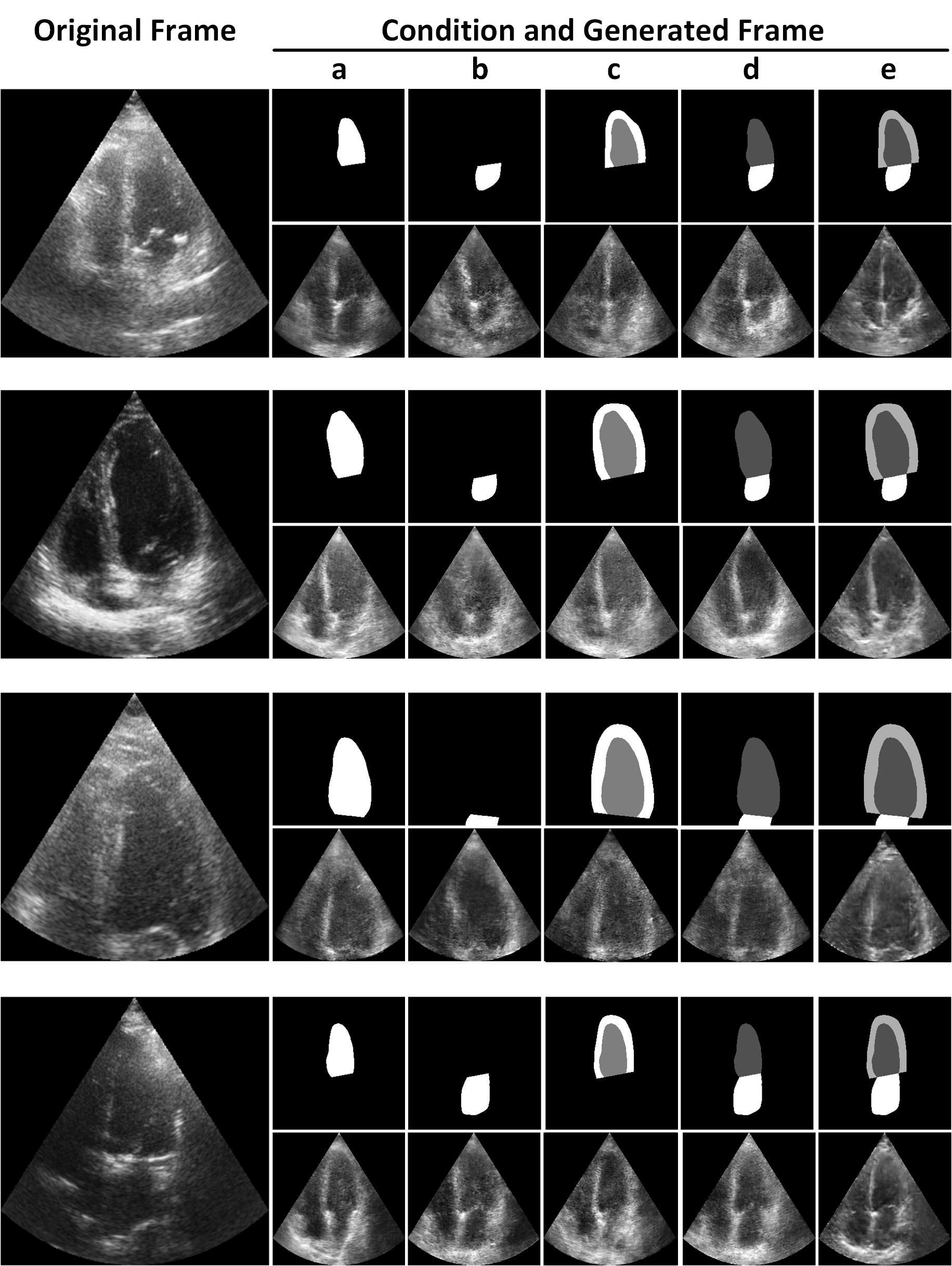}
\caption{More test results demonstrating models' performance dealing with  challenging cases; a) Left ventricle b) Left atrium, c) Left ventricle and ventricular myocardium, d) Left ventricle and atrium, e) Left ventricle, ventricular myocardium, and atrium.} 
\label{fig:results2}
\end{figure}

The generative model is blind to the fact that the original echo image, from which the segmentation mask (condition) is extracted, may not have captured the cardiac structures  clearly.
Therefore, and regardless of the original echo, the generated echo is always of high quality.
For example, in the last row of Figure~\ref{fig:results2}, even though the original frame has  sub-optimal quality, the generated frames do not suffer from the same problem.

\begin{figure}[!h]
\includegraphics[trim={0 0 0 0},clip ,width=1.0\textwidth]{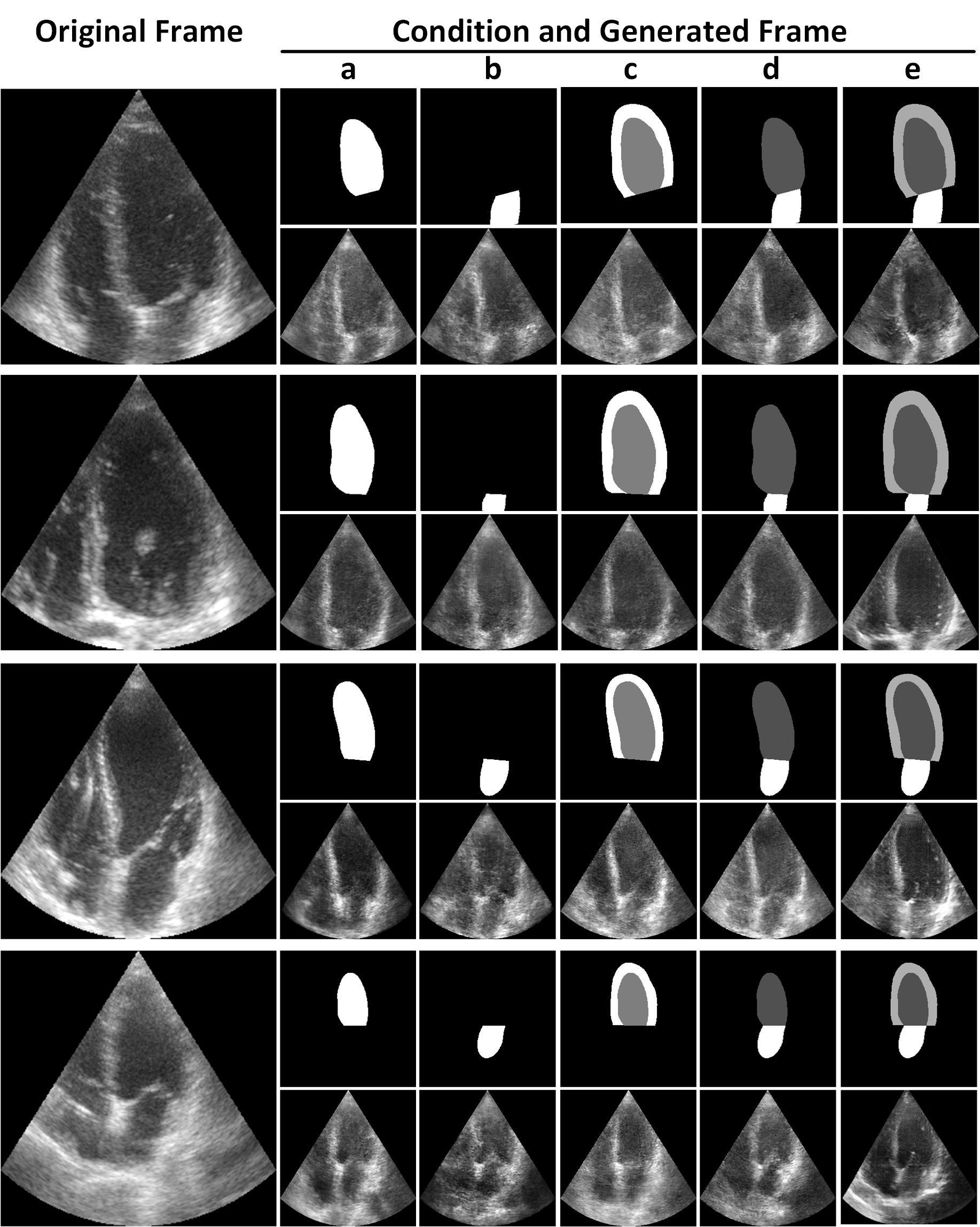}
\caption{Generated end-diastolic frames conditioned on  the segmentation masks  along with their corresponding original (real) frames; a) Left ventricle b) Left atrium, c) Left ventricle and ventricular myocardium, d) Left ventricle and atrium, e) Left ventricle, ventricular myocardium, and atrium.} 
\label{fig:results1}
\end{figure}

\end{document}